\documentclass[onecolumn,aps,prc,nofootinbib,floatfix]{revtex4}

\usepackage{amsmath}
\usepackage{bm}
\usepackage{graphicx}

\usepackage{epsfig}
\newcommand{\beq}{\begin{equation}}
\newcommand{\eeq}{\end{equation}}
\newcommand{\bea}{\vspace{0.25cm}\begin{eqnarray}}
\newcommand{\eea}{\end{eqnarray}}

\newcommand{\ro}{\mbox{{\boldmath
$\rho$}}}

\def\lsim{\mathrel{\rlap{\lower4pt\hbox{\hskip1pt$\sim$}}
    \raise1pt\hbox{$<$}}}         
\def\gsim{\mathrel{\rlap{\lower4pt\hbox{\hskip1pt$\sim$}}
    \raise1pt\hbox{$>$}}}         

\newcommand{\landau}{L.D.~Landau Institute for Theoretical Physics,
        GSP-1, 117940, Kosygina Str. 2, 117334 Moscow, Russia}

\begin{document}


\title{
Radiative parton energy loss and baryon stopping in $AA$ collisions
}
\date{\today}

\author{B.G.~Zakharov}\affiliation{\landau}

\begin{abstract}
We study the radiative energy loss contribution to  
proton stopping in $AA$ collisions.
The analyses is performed within the light-cone path integral approach
to induced gluon emission.
We have found that  the radiative correction  can fill in partly the 
midrapidity dip in the net proton rapidity distribution in $AA$ collisions at
$\sqrt{s}\sim 10$ GeV. 
We argue that at $\sqrt{s}\sim 10$ GeV  the net proton fluctuations 
at midrapidity may be dominated by the initial fluctuations of 
the proton flow, which, to a good accuracy, should be
binomial. 

\end{abstract}
%

\maketitle
\noindent {\bf 1}.
The baryon stopping in hadron and nucleus collisions has attracted 
much attention for a long time. 
But up to now, there is no consensus yet on the mechanism of the baryon number
transfer over a large rapidity interval (which is also closely related to 
the mechanism of $B\bar{B}$-annihilation). Presently, there is no 
answer to the most basic question about the baryon production:
whether the baryon number carriers are quarks. 
It was proposed long ago \cite{RV1} that a purely gluonic
object--the so-called string junction (SJ) may play role of the 
baryon number carrier.
In \cite{RV1} it was suggested to describe the processes with baryons in  
the topological expansion (TE) scheme \cite{Ven_TE,Ven_TEQCD}
by treating the baryon as a Y shaped string configuration with the SJ. 
In this picture the baryon number transfer is associated with 
the transfer of the SJ. Say,  $B\bar{B}$ annihilation corresponds to 
SJ-$\overline{\text{SJ}}$ annihilation. 

In the standard quark-gluon string model QGSM \cite{DTU_DPM,DPM1,QGSM_K},
based on the TE scheme \cite{Ven_TE},
it is assumed that the baryon can be treated as a quark-diquark system
with a point like diquark in the $\{\bar{3}\}$ color state.
From the point of view of the SJ picture \cite{RV1}, 
it means that the SJ belongs to diquark (we denote this mechanism by D-SJ).
In this approximation diquarks/antidiquarks are playing the role of the baryon number 
carriers in processes with baryons. 
In the QGSM the net proton ($\Delta p=p-\bar{p}$) 
midrapidity density in $pp$ collisions is related to the diquark distribution
in the proton at small fractional momentum $x$, where
it is $\propto x^{\alpha_R(0)-2\alpha_N(0)}$ with 
$\alpha_R(0)\approx 0.5$ and $\alpha_N(0)\approx-0.5$ intercepts of the
meson and nucleon Regge trajectories \cite{KP}.
This leads to the energy dependence of the net proton
midrapidity density $dN_{\Delta p}/dy\propto 1/s^{2.25}$. This disagrees
strongly with the experimentally observed at ISR energies \cite{ISR}
$s$-dependence $dN_{\Delta p}/dy\propto 1/s^{n}$ with $n\approx 0.25$.
The model with the diquarks in the antitriplet color state also
underestimates the midrapidity net baryon production in $AA$ collisions
\cite{KC-AA,C-AA99}. 

In \cite{Z_B1} it was proposed that the baryon number transfer over a large
rapidity interval can be related to breaking 
of the antitriplet diquark due to its transition
from $\{\bar{3}\}$ to $\{6\}$ color state after one gluon
$t$-channel exchange. The transition $D_{\{\bar{3}\}}\to D_{\{6\}}$ 
should lead to creation of the string configuration with the SJ
located near the valence quark (we denote this mechanism as q-SJ).  
Hadronization of such string configurations
leads naturally to the energy dependence of the midrapidity net proton density
$dN/dy\propto 1/s^{\alpha_R(0)/2}$ which agrees with the data.
Contribution of the q-SJ mechanism of the baryon number transfer 
over a large rapidity interval should be enhanced in $AA$ collisions 
due to increase of the probability of the $D_{\{\bar{3}\}}\to D_{\{6\}}$ 
transition in the nuclear matter. Similarly to $pp$ collisions, in 
$AA$ collisions the diquark breaking leads to the net proton
midrapidity distribution $\propto 1/s^{\alpha_R(0)/2}$ \cite{C-AA96}.

The contribution from the q-SJ mechanism to the net proton midrapidity
density in $AA$ collisions becomes of the order of that
from the ordinary D-SJ mechanism at $\sqrt{s}\sim 20$ GeV.
At energies $\sqrt{s}\sim 10$ GeV contribution
of the q-SJ mechanism is relatively small.
One more effect of the nuclear matter that can increase the
baryon number flow to the midrapidity,
which  can potentially  be important at $\sqrt{s}\lsim 10$ GeV, is the diquark 
radiative energy loss.
At high energies the radiative energy loss (which is not very large) 
cannot increase considerably the contribution of the D-SJ 
mechanism, and the radiative correction to the D-SJ mechanism  cannot compete 
with the q-SJ mechanism.
But at $\sqrt{s}\lsim 10$ GeV, where
the contribution of the q-SJ mechanism becomes small,
the radiative mechanism might be important, and it is of great interest 
to understand how large the radiative contribution to the baryon 
stopping can be.

The question of how the baryon number flows to the midrapidity,
which is interesting in itself, 
at energies $\sqrt{s}\lsim 10$ GeV is also of great importance in connection
with the future experiments at NICA and 
the beam energy scan (BES) program at RHIC
aimed to search for the QCD critical point. 
One of the important signals of the critical point may be the non-monotonic
variation of the moments of the net-baryon distribution
in the central rapidity region \cite{Steph1}. 
However, in the event-by-event fluctuations
of the net-baryon yield a considerable contribution
may come from the initial state fluctuations \cite{Asakawa2}. 
One can expect that, similarly to the situation with the baryon number 
fluctuations in the QGP and the hadron gas \cite{Asakawa1,Asakawa2}, for 
the D-SJ mechanism the  variance for the net-baryon number 
should be bigger by a factor of $\sim 3$ than in the picture 
with the baryon number associated with quarks.

In the present paper we calculate the the diquark 
and quark radiative energy loss in $AA$ collisions within the light-cone path integral (LCPI) 
approach \cite{LCPI1,LCPI2,LCPI3} to the induced gluon emission. We study
the effect of the diquark radiative energy loss in the nuclear matter 
on the net proton rapidity distribution for the D-SJ mechanism.
We perform comparison with the data on the proton distribution
in Pb+Pb collisions at $\sqrt{s}=8.76$ GeV from NA49 Collaboration
\cite{NA49}. \\

\noindent {\bf 2}.
The scalar $ud$ diquark, which  dominates in the nucleon wave function
\cite{Ans,D1,D2}, is a rather compact object. It has a formfactor
$F(Q^2)\approx 1/[1+Q^2/Q^2_0]$ with $Q^{2}_0\approx 10$ GeV$^2$ \cite{Ans}.
The typical virtuality scale for the induced gluon emission
in the nuclear matter ($\sim m_{g}^2\lsim 0.5$ GeV$^2$) 
is much smaller than $Q^{2}_{0}$. 
For this reason the induced gluon radiation from scalar diquarks
can be calculated as for a point like particle. This approximation
should be reasonable even for the less compact vector $ud$ and $uu$ 
diquarks, for which 
$Q^2_0\approx 2$ GeV$^2$ \cite{Ans}.

We will treat nuclei as uniform spheres. We consider
$AA$ collisions in the rest frame of one (target) of the colliding nuclei.
To a good approximation, 
the induced gluon emission from each diquark/quark in the projectile
nucleus can be evaluated similarly to $pA$ collisions, i.e.
ignoring the presence of other nucleons.
Indeed, the typical transverse size for the induced gluon emission
is $\sim 1/m_g$. The typical number of nucleons in the tube 
of radius $r\sim 1/m_g$ in the projectile nucleus
is $\mu \sim A/{R_{A}^2m_g^2}$. For $m_g\sim 400-800$ MeV
it gives $\mu \sim 0.4-1.5$ for heavy nuclei with $A\sim 200$.
This estimate shows that the possible collective effects in
the projectile nucleus should not be strong.

In the LCPI approach the $x$-spectrum of the induced $a\to bc$ transition
in the fractional longitudinal momentum $x=x_b=E_b/E_a$ can be written as
(we take the $z$-axis along the projectile momentum) \cite{LCPI1,LCPI3}
\beq
\frac{d P}{d x}=2\mbox{Re}\!
\int_{-\infty}^{\infty} d z_{1}\int_{z_1}^{\infty} d z_2
\exp\left[-{i(z_{2}-z_{1})}/{L_{f}}\right]
\hat{g}\left[{\cal{K}}(\ro_2,z_{2}|\ro_1,z_{1})
-{\cal{K}}_{0}(\ro_2,z_{2}|\ro_1,z_{1})\right]\Bigg|_{\ro_{1,2}=0}\,.
\label{eq:10}
\eeq
Here ${\cal{K}}$ is the Green's function for the 
Schr\"odinger equation with 
the Hamiltonian 
\beq
H=-\frac{1}{2M}\,
\left(\frac{\partial}{\partial \ro}\right)^{2}
+v(\ro,z)\,,
\label{eq:20}
\eeq
\beq
v(\ro,z)=-i\frac{n(z)\sigma_{3}(\rho,x)}{2}\,,
\label{eq:30}
\eeq
where 
$
M(x)=E_{a}x(1-x)\,
$
$
L_{f}=2M/[m_{b}^{2}x_c+m_c^2x_b-m_{a}^{2}x_bx_c]
$, $n$ is the medium density, and
$\sigma_{3}(\rho,x)$ is the cross section of interaction
of the $bc\bar{a}$ system with a nucleon,
${\cal{K}}_0$ is the vacuum Green's function for $v=0$.
$\hat{g}$ is the vertex factor given by
\beq
\hat{g}=\frac{\alpha_{s}P_{ba}}{2 M^2}\cdot
\frac{\partial}{\partial \ro_1}
\cdot\frac{\partial}{\partial \ro_2}\,,
\label{eq:40}
\eeq
where $P_{ba}$ is ordinary splitting function 
(for $q\to gq$ $P_{gq}=C_F[1+(1-x)^2]/x$, and 
for $D\to gD$ $P_{gD}=2C_F(1-x)/x$).
The three-body cross section, 
entering the imaginary potential
(\ref{eq:30}), for the $D(q)\to gD(q)$ processes
can be expressed
in terms of the dipole cross section for the color singlet $q\bar{q}$
pair \cite{SIG3}
\beq
\sigma_{3}(\rho,x)=\frac{9}{8}[\sigma_{2}(\rho)
+\sigma_{2}((1-x)\rho))]-\frac{1}{8}\sigma_{2}(x\rho)\,.
\label{eq:50}
\eeq

In the limit of $L_f\gg L$  the radiation rate is dominated by the
configurations with large negative $z_1$ and large positive $z_2$
$|z_{1,2}|\gg L$ ($L$ is thickness of the target, 
for a nucleus with radius $R_A$ $\langle L\rangle\approx 4R_A/3$). In this
regime the spectrum (\ref{eq:10}) may be calculated
treating the transverse parton positions 
during traversing the target as frozen.
The integrals over $z_{1,2}$ outside the target region on the right hand side of (\ref{eq:10}) 
can be expressed via the vacuum light-cone wave function for the $a\to bc$
\cite{LCPI3}, and the Green' function in the medium is reduced 
to the ordinary Glauber attenuation factor for the $bc\bar{a}$ system.
This leads to the formula \cite{NPZ,LCPI3}  
\beq
\frac{dP_{fr}}{dx}=
2\int d\ro
|\Psi(x,\ro)|^2
\left\{1-\exp\left[-\frac{nL\sigma_3(\rho,x)}{2}\right]\right\}\,,
\label{eq:60}
\eeq
where $\Psi$ is the light-cone wave function of the $bc$ system in the
$(x,\ro)$ representation.
The  frozen-size approximation is widely used in physics of $AA$ collisions,
e.g., it was used for evaluation of the gluon spectrum in \cite{MK}.
This approximation should be good for emission of gluons with 
energies $\omega \gg 10-20$ GeV (in the target nucleus rest frame). But for $AA$ collision at 
$\sqrt{s}\sim 10$ GeV, interesting to us here, 
the typical gluon energies turn out to not big enough 
for  its applicability.

We will perform calculation for the quadratic 
approximation $\sigma_{q\bar{q}}(\rho)=C_2\rho^2$
($C_2$ may be written in terms
of the well known transport coefficient
$\hat{q}$ \cite{BDMPS} as $C_2=\hat{q}C_F/2nC_{A}$).
In this case the Hamiltonian (\ref{eq:20}) in the medium 
takes an oscillator form
with the complex frequency given by 
$
\Omega=\sqrt{-inC_3/M}\,
$
with $C_3=C_2\{\frac{9}{8}[1
+(1-x)^2]-\frac{x^2}{8}\}$.
In the oscillator approximation 
the Green's function can be written 
in the form
\beq
{\cal{K}}(\ro_2,z_2|\ro_1,z_1)=\frac{\gamma}{2\pi i}
\exp{\left[i(\alpha\ro_2^2+\beta \ro_1^2-\gamma\ro_1\cdot\ro_2)\right]}\,.
\label{eq:70}
\eeq
Only the parameter $\gamma$ is important in calculating the $x$-spectrum.
From (\ref{eq:10}), (\ref{eq:70}) one can obtain 
\beq
\frac{d P}{d x}=\frac{\alpha_sP_{ba}(x)}{\pi M^2}
\mbox{Re}
\int_{-\infty}^{\infty} d z_{1}\int_{z_1}^{\infty} d z_2
\exp\left[-{i(\xi_{2}-\xi_{1})}/{L_{f}}\right]
\cdot(\gamma_0^2-\gamma^2)\,.
\label{eq:80}
\eeq
Here $\gamma_0=M/(z_2-z_1)$ is the parameter $\gamma$ in
(\ref{eq:70}) for vacuum when $\Omega=0$, and 
\beq
\gamma=M\Omega\times
\left\{
\begin{array}{ll}
{\phantom{\Big |}\sin^{-1}{(\Omega(z_2-z_1))}}&
\text{at }\,\,\,L>z_2>z_1>0\,,\\
\phantom{\Big |}\{\cos{(\Omega(L-z_1))}\cdot[\tan{(\Omega(L-z_1))}+\Omega(z_2-L)]\}^{-1} &
  \,\,\,\,\text{at}\,\,
z_2>L>z_1>0\,,\\
\phantom{\Big |}\{\cos{(\Omega z_2)}\cdot [\tan{(\Omega z_2)}+\Omega |z_1|)]\}^{-1} &
  \,\,\,\,\text{at}\,\,
L>z_2>0> z_1\,,\\
\phantom{\Big |}\{\cos{(\Omega L)}\cdot[\Omega(z_2-z_1-L)
+\tan{(\Omega L)}(1-\Omega^2(z_2-L)|z_1|)]\}^{-1} & \text{at}\,\, z_2>L,\,\,z_1<0\,.\\
\end{array} \right.
\label{eq:90}   
\eeq\\

\noindent {\bf 3}.
For numerical calculations we take $\alpha_s=0.5$.  
We use the value $\hat{q}=0.01$ GeV$^3$, supported by
calculations of the coefficient $C_2$ using the double gluon formula
for the dipole cross section \cite{LCPI2}. 
For the quark and diquark 
masses we take $m_q=m_D=300$ MeV. However, the radiative energy loss 
is only weakly dependent on the specific choice of the mass of the initial
fast particle. But its dependence on the gluon mass, which plays the role
of the infrared cutoff, turns out to be rather
strong \footnote{Note the the situation with the infrared sensitivity of
the radiative energy loss for a parton incident on the medium from outside
is very different from that for a parton produced inside the medium.
In the former case the fast parton approaches the target
with a formed gluon cloud with the transverse size $\sim 1/m_g$.
While in the latter case the fast parton is produced without the 
formed gluon cloud. After gluon emission the transverse size of
the two-parton system grows $\propto \sqrt{L/\omega}$ \cite{Z_OA}. 
As a result
for gluons with $L_f\gsim L$ the spectrum depends weakly on the
gluon mass. This leads to a weak infrared sensitivity
of the total parton energy loss \cite{Z_OA}.}.
We perform numerical computations for $m_g=750$ MeV 
and $m_g=400$ MeV.
The former value was obtained in the analysis 
within  the  dipole  BFKL  equation  \cite{NZ_HERA}
of  the  data on the low-$x$
proton  structure  function $F_2$.
The values of $m_g$  in the range $400-800$ MeV have been predicted
in the nonperturbative calculations using the 
Dyson-Schwinger equations \cite{DSE_m04,DSE_m05,DSE_m08}.
We take for Pb nucleus $R_A=6.49$ fm, that corresponds to the typical parton
path length in the nuclear matter $\langle L\rangle\approx 8.65$ fm.

In Fig.~1 we present the result for 
the total radiative
energy loss for diquarks and quarks for Pb nucleus 
\beq
\Delta E=E\int_{x_{min}}^{x_{max}} dx x \frac{dP}{dx}\,.
\label{eq:100}
\eeq
We take $x_{min}=m_g/E$ and $x_{max}=1-m_q/E$. To illustrate 
the effect of the parton transverse motion we show in Fig.~1 also
the prediction obtained with the spectrum calculated 
in the frozen-size approximation (\ref{eq:60}). One can see that
the frozen-size approximation underestimates the energy loss by 
a factor of $1.6-1.7$($1.2-1.3$) at $m_g=400$($750$) MeV.
Thus, the frozen-size approximation turns out to be rather crude.
Fig.~1 shows that the ratio $\Delta E/E$ flatten at $E\gsim 50-100$ GeV,
and at $E\sim 1000$ GeV for diquark $\Delta E/E\approx 0.18(0.062)$ 
at $m_g=400$($750$) MeV. Note that some violation of the $1/m_g^2$ scaling
for $\Delta E$
is due to the Landau-Pomeranchuk-Migdal suppression that is stronger
for smaller $m_g$.

For evaluation of the net baryon spectrum within the QGSM one needs only the
contribution of the diquark fragmentations to baryons, i.e. the contribution 
from color strings with the valence diquarks. 
In $pp$ and $AA$ collisions the contributions from the projectile nucleons (diquark-quark strings) and from
the target nucleons (quark-diquarks strings) can be evaluated independently.
The valence quarks and sea quarks/diquarks are irrelevant. 
The net baryon $x$-distribution for each projectile nucleon 
in the QGSM can be written as \cite{KC-AA,C-AA99}
\beq
\frac{dN_{AA}^{\Delta B}}{dx}\approx \int_{x}^1 \frac{dz}{z}
\rho_D(z)D_D^B(x/z)\,,
\label{eq:110}
\eeq
where $\rho_D$ is the diquark distribution in the 
projectile nucleon, $D_D^B$ is the $D\to B$  fragmentation function.
We use for the diquark distribution the parametrization 
of the form used in the \cite{KC-AA,C-AA99}
\beq
\rho_D(x)=C x^{\alpha_R(0)-2\alpha_N(0)}(1-x)^{-\alpha_R(0)-1+k}\,,
\label{eq:120}
\eeq
where $C=[\int_0^1 dx \rho_D(x)]^{-1}$ is the normalization constant.
The parameter $k$ controls the energy degradation due to creation
of additional sea $q\bar{q}$ pairs in the nucleon, that lead 
to formation of additional color strings in the final state. 
Within the QGSM there is no a rigorous  theoretical method for its 
determination. It is input by hand in order to reproduce 
experimental data. In the case of $pp$ collisions data on the charged 
particle multiplicities can be reasonably described using the quasieikonal 
formulas \cite{C-ppAA12}.
However, the model fails to describe the $AA$ data \cite{C-ppAA12}.
In \cite{KC-AA,C-AA96, C-AA99} the value of $k$ was set 
to $\nu=2N_{col}/N_{part}$ (here $N_{col}$ and $N_{part}$ are the number of the
binary collisions and the number of participant nucleons
evaluated in the Glaber model). In terms of the wounded nucleon model
\cite{WNG,KN} $\nu$ is simply the average number of inelastic collisions
per participant nucleon. This prescription gives reasonable agreement
with the data on charged particle multiplicity in $AA$ collisions
\cite{KC-AA}. In the present analysis we use a slightly modified formula
$k=\nu+\langle k_N\rangle -1$, where $\langle k_N\rangle$ is the average 
number of cut Pomerons in the quasieikonal formulas for  $pp$ collisions
($\langle k_N\rangle\approx 1.65$ at $E=40$ GeV).
This modification practically does not change the charged particle 
multiplicity for $AA$ collisions, but it somewhat improves description 
of the data on the baryon stopping in $pp$ and $AA$ collisions.
Also, this prescription ensures matching of the predictions for very peripheral $AA$ collisions with that for $pp$ collisions.

The induced gluon emission in $AA$ collisions softens the diquark  
and quark distributions due to the radiative energy shift. 
However, if one neglects the diquark transitions to 
the $\{6\}$ color state, 
the radiative change of the quark 
distribution is irrelevant for the baryon stopping.
But the radiative diquark energy loss should increase the baryon
stopping as compared to the predictions of the QGSM. 
In the presence of the induced gluon emission we write the diquark
distribution 
as (we denote it by $\rho_D^{eff}$)
\beq
\rho_D^{eff}(x)=\rho_D(x)+\Delta \rho_D(x)\,,
\label{eq:130}
\eeq
where $\Delta\rho_D(x)$ is the radiative correction.
In terms of the induced gluon spectrum (\ref{eq:10}), the radiative correction
to the QGSM diquark distribution $\rho_{D}$ can 
be written as
\beq
\Delta \rho_D(x)=\int_{x_{min}}^{1-x} dz \frac{dP}{dz}\left[\frac{\rho_D(x/(1-z))}{1-z}-
\rho_D(x)\right]\,.
\eeq
Here the second term in the square brackets is due to reduction
of the probability to find the diquark without gluon emission.
Note that due to this term the radiative softening of the diquark distribution
is insensitive to the $z$-dependence of the gluon spectrum at $z\to 0$.
\begin{figure}
\includegraphics[width=8cm]{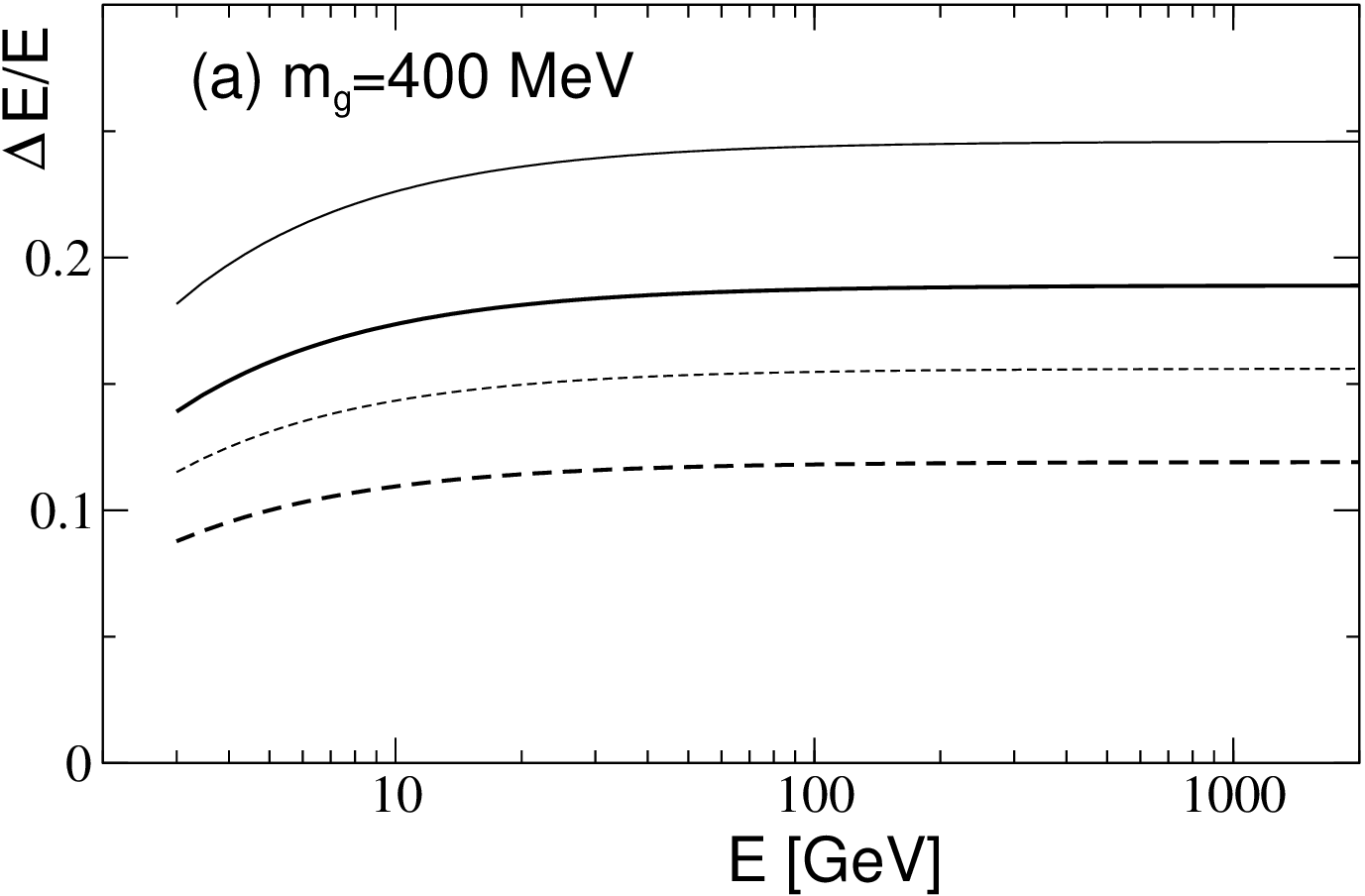}\hspace{.5cm}
\includegraphics[width=8cm]{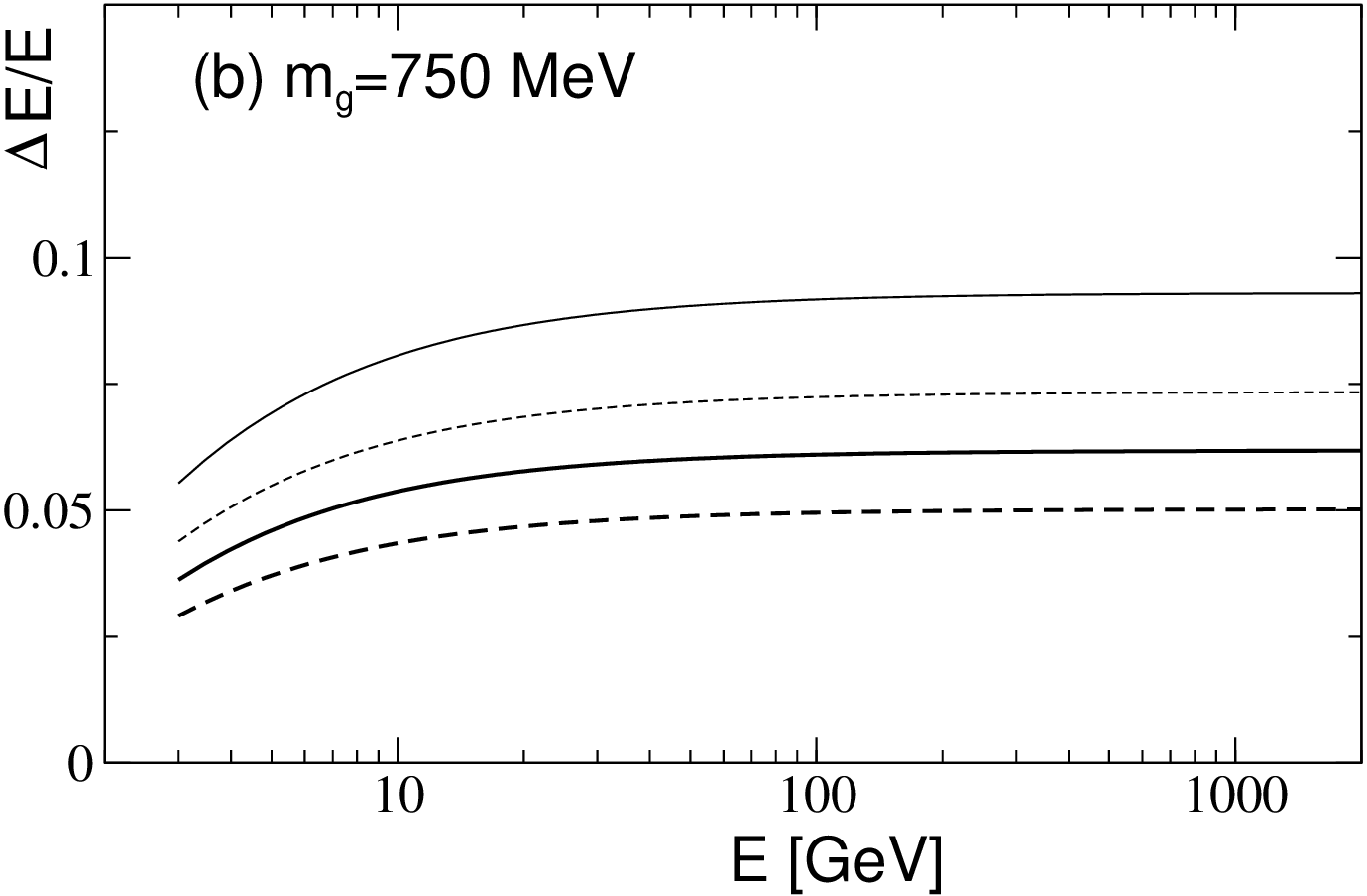}
\caption{\small The fractional 
diquark (thick solid) and 
quark (thin solid) radiative energy loss versus energy for Pb nucleus
obtained using (\ref{eq:100}) with 
accurate gluon spectrum (\ref{eq:80}) for (a) 
$m_g=400$ MeV and (b)
$m_g=750$ MeV. The dashed line shows same
for gluon spectrum calculated in the frozen-size approximation
(\ref{eq:60}).
}
\end{figure}
\begin{figure}
\includegraphics[width=8cm]{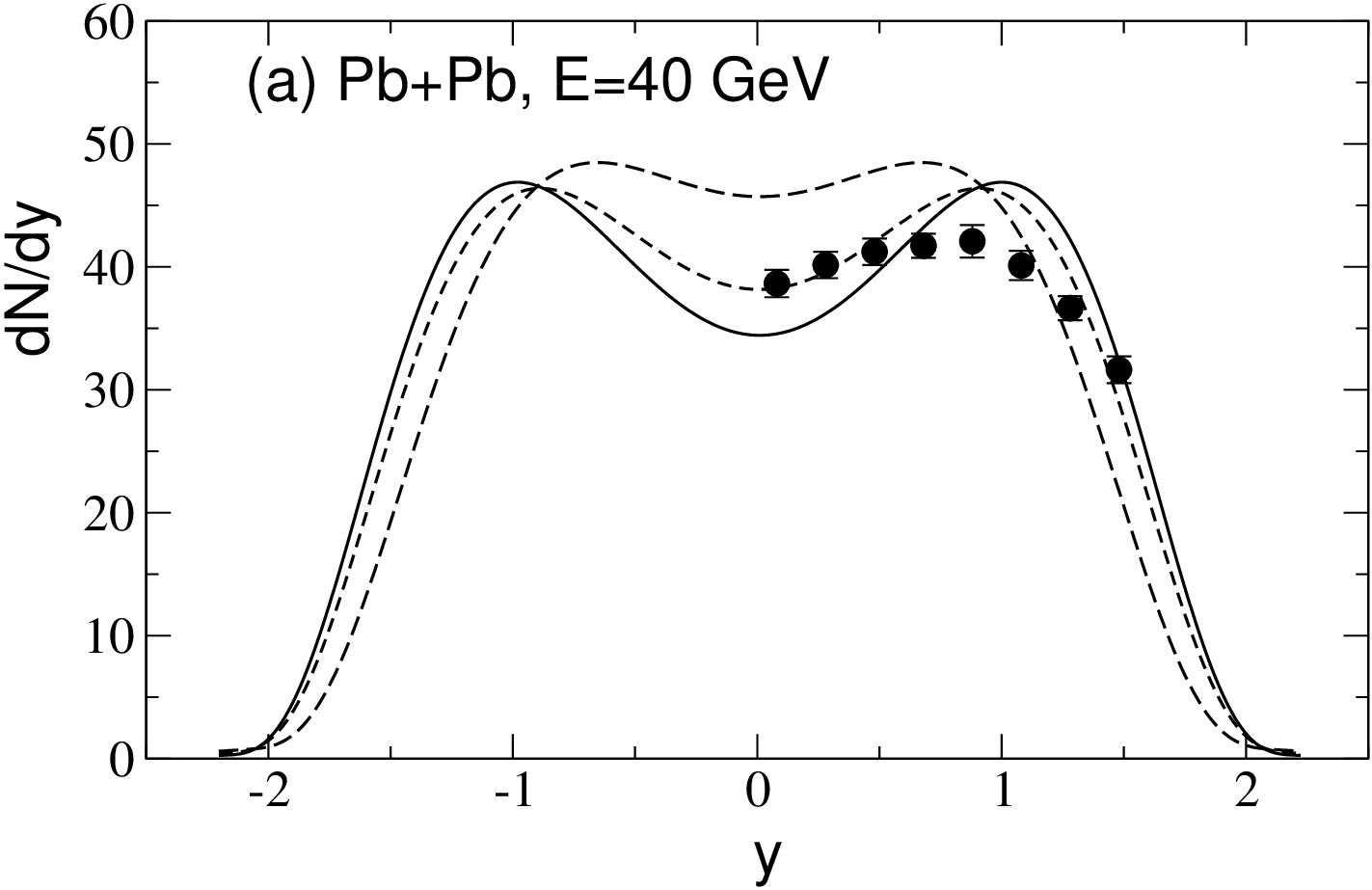}\hspace{.5cm}
\includegraphics[width=8cm]{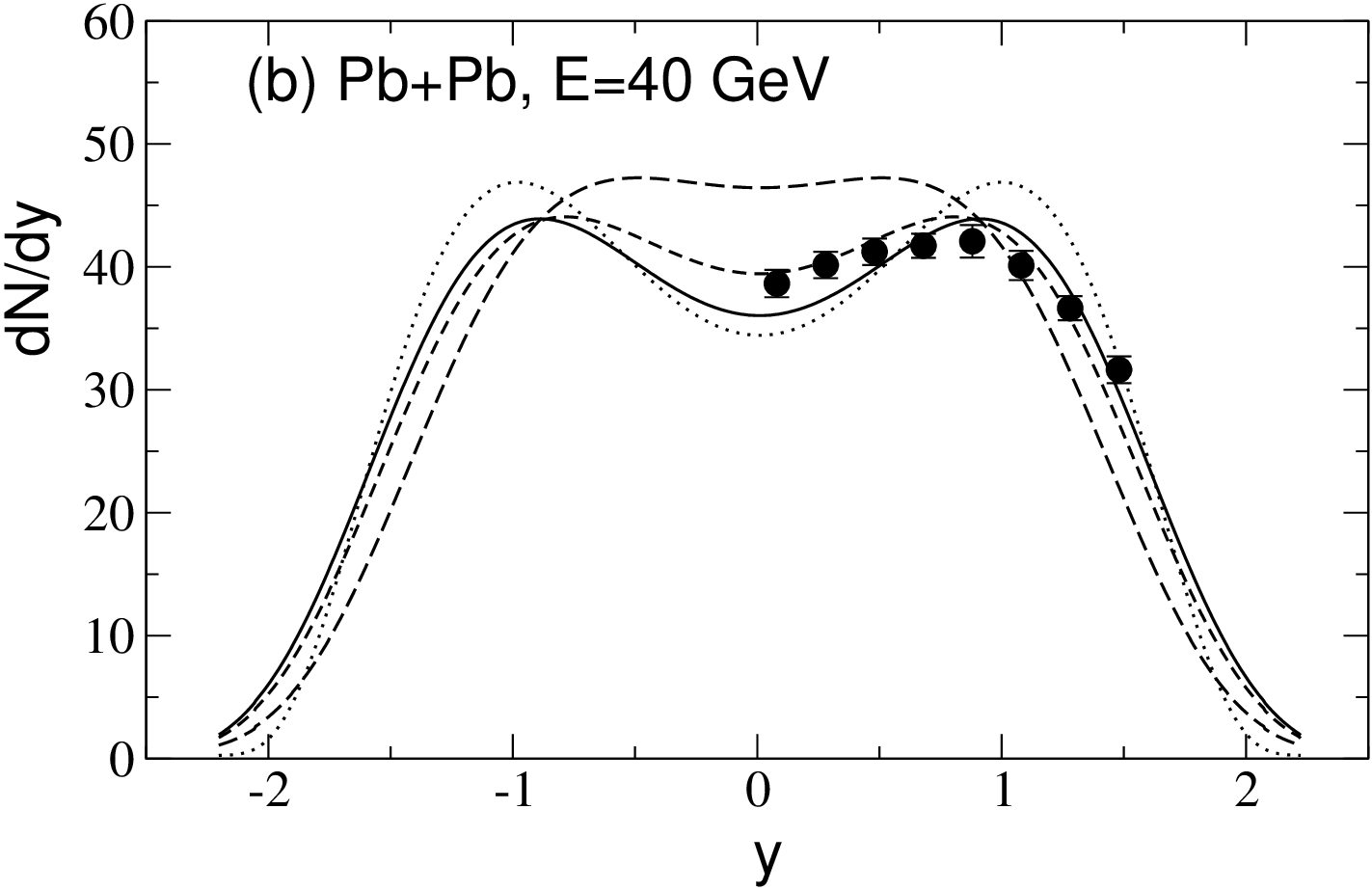}
\caption{\small (a) The net proton rapidity distribution in $0-5$\% central 
Pb+Pb collisions
at $E=40$ GeV obtained without (solid) and with the radiative correction
for $m_g=400$ MeV (long dashes) and for $m_g=750$ MeV (short dashes).
Data points are from NA49 \cite{NA49}.
(b) Same as (a) accounting for the Gaussian smearing with the diffusion
width $\sigma_d=0.3$, the dotted line shows the distribution without
radiative and diffusion corrections (same as solid line in the left panel).
}
\end{figure}

In Fig.~2a we confront the results of our calculations with and without the induced gluon emission  of 
the net proton rapidity spectrum (in the center of mass frame) 
with the data from NA49 Collaboration
for  Pb+Pb collisions at $E=40$ GeV ($\sqrt{s}=8.76$ GeV) \cite{NA49}.
The theoretical curves are obtained by summing the contributions of the
valence proton flow from the projectile and the target nuclei.
The calculations are performed with the fragmentation function $D\to p$ from
\cite{CK-Bflow} $D_D^p(z)=a z^{1.5}$. The value of the normalization
parameter $a$ depends on the strange suppression factor $S/L$ ($2L+S=1$) introduced 
in \cite{CK-Bflow}, for which we take $S/L=0.367$. This value was adjusted to reproduce 
the experimental midrapidity ratio $p/\Lambda\approx 0.4$ for Pb+Pb collisions
at $E=40$ GeV \cite{NA49,NA49Lambda}.
From Fig.~2a one sees that the radiative correction 
partly fills in the minimum at $y=0$. It 
increases the spectrum at $y=0$ by a factor
of about $1.35$ and $1.12$ for $m_g=400$ and $750$ MeV, respectively.
Thus, we conclude that for $m_g=750$ MeV the radiative effect is relatively 
weak. For $m_g=400$ MeV the radiative effect is quite strong,
and the theoretical spectrum overshoot the data at $y\sim 0$.
Of course, one can improve agreement with the data
by changing the QGSM parton distributions, which
have not stringent theoretical constraints in the QGSM.
However, in this preliminary study our purpose
is to understand if the radiative energy loss may be important, and
not as good as possible fitting to the data.

Our results shown in Fig.~2a do not include the possible baryon diffusion
in the hot QCD matter produced after interaction of the colliding nuclei at the proper
time $\tau_0\approx 2R_A/\gamma$ ($\gamma$ is the nucleus Lorentz factor in the
center of mass frame). For $E=40$ GeV ($\gamma\approx 4.7$) we have
$\tau_0\approx 2.8$ fm. The change of the baryon rapidity is related 
to its random walk in the longitudinal $z$ direction  in the comoving  frame 
of the QCD matter. For the diffusion mean squared rapidity fluctuations
one can easily obtain
\beq
\langle \Delta y_d(\tau)^2\rangle^{1/2}\approx \int_{\tau_0}^{\tau}
\frac{\langle \Delta z(\tau)^2\rangle^{1/2}}{d\tau}\frac{d\tau}{\tau}\,.
\label{eq:150}
\eeq 
Making use of the random walk formula 
$\langle \Delta z(\tau)^2\rangle=(\tau-\tau_0) \bar{v}l_c/3$
(here $\bar{v}$ is the mean proton velocity, $l_c$ is the proton mean free path length) 
from (\ref{eq:150}) one obtains 
\beq
\langle \Delta y_d(\tau)^2\rangle^{1/2}\approx \sqrt{\frac{\bar{v} l_c}{3\tau_0}}
\left[\frac{\pi}{2}-\arcsin\left(\sqrt{\tau_0/\tau_f}\right)\right]\,,
\label{eq:160}
\eeq
where $\tau_f$ is the freeze out proper time.
From the relation for the initial entropy density 
$
s_{0}=\frac{C}{\tau_{0}\pi R_{A}^{2}}\frac{dN_{ch}}{d\eta}\,
$
($C=dS/dy{\Big/}dN_{ch}/d\eta\approx 7.67$ 
is the entropy/multiplicity ratio \cite{BM-entropy}),
using the midrapidity charged particle density $dN_{ch}/d\eta\approx 280$ 
\cite{NA49Nch,NA49} and the lattice QCD matter EoS \cite{entropy}
we obtain $T_0\approx 170-175$ MeV for $\tau_0\approx 2.8 $ fm.
Using the hadron gas model for the relevant range  
of the temperature $T\sim 140-175$ MeV
from (\ref{eq:160}) with $\tau_f\approx 13$ fm \cite{FROUT}
one obtains  
$\langle \Delta y_d(\tau_f)^2\rangle^{1/2}\approx 0.3$.
To illustrate the magnitude of the diffusion correction,
in Fig.~2b we show the prediction obtained with the Gaussian
smearing with the width $\sigma_d=0.3$.
One sees that the smearing partly fills in
the minimum.

From Figs.~2a,~b one can see that the diffusion effect 
should not affect strongly the net proton rapidity distribution.
However, one can expect that for $\sigma_d\sim 0.3$ 
the diffusion may be important for 
the event-by-event net proton fluctuations for the rapidity 
window $\Delta y$ about $\sim 2-3$
units of $\sigma_d$. In this regime the diffusion can 
modify somewhat the primordial net proton yield fluctuations.
But, at the same time, it is clear that for $\Delta y/\sigma_d\sim 3$ 
this modification cannot be strong. For this reason the net baryon 
charge fluctuations cannot be described by the grand canonical
ensemble formulas which require $\sigma_d\gg \Delta y$,
and one cannot expect to observe a real critical regime.
In the light of this, the absence of a clear signal 
of the critical point in the net proton fluctuations
at $\sqrt{s}\sim 10$ GeV for $|y|<0.5$ in Au+Au collisions  
from STAR \cite{STAR_BES} is not surprising. 
For the diquark mechanism of the baryon flow, to a good accuracy, 
these fluctuations should be binomial. 
This agrees with the STAR observation \cite{STAR_BES}
that the net proton fluctuations are close to binomial/Poissonian 
at $\sqrt{s}\sim 10$ GeV, where the antiproton yield becomes very small.
In the scenario of \cite{Kovch},
in which the carriers of baryon number are quarks,
the binomial/Poissonian distribution occurs for the net quark fluctuations.
In this case 
$\langle (N_{\Delta p}-\langle N_{\Delta p}\rangle)^2\rangle\approx
\langle N_{\Delta p}\rangle/3$ (cf. \cite{Asakawa1}),
which contradicts the STAR measurement \cite{STAR_BES}.
Note that even without calculating the diffusion width $\sigma_d$,
the existence of the dip in the experimental net proton
distribution at $y=0$ from NA49 \cite{NA49} says that
$\sigma_d$ is considerably smaller than $\sim 1$.
Because, numerical calculations show that 
for $\sigma_d\sim 1$ the diffusion
should completely wash out the dip. So, one can say, that we have an
experimental evidence that 
at $\sqrt{s}\sim 10$ GeV the net proton fluctuations for 
$\Delta y\sim 1$ cannot not be close the critical 
point regime.

One remark is in order regarding the status of the parton radiative energy loss
in the nuclear matter in the string model. One might wonder whether the adding
of the radiative energy loss leads to the double counting, because the QGSM already includes the
energy degradation of the ends of the strings for multiple Pomeron exchanges.
However, in the TE/QGSM the energy degradation is due to creation of the 
parallel hadronic states, say, $h_1\to h_2h_3$ for two Pomeron exchanges. 
In the TE \cite{Ven_TE}, based on the $1/N$ expansion 
($N=N_c$, $N_F/N_c=$const), such parallel energy degradation
is due to the quark loops that survive at $N\to \infty$, only 
because $N_F\to \infty$ together with  $N_c$. On the contrary, 
the induced gluon
emission is a purely gluonic effect, which does not depend on $N_F$ at all. 
Contrary to the parallel energy degradation in the TE, this mechanism 
does not increase the number of the color
strings (as in the LUND model \cite{LUND} the radiated gluon 
may be treated as a kink on the available string attached to the diquark). 
Thus, it is clear the radiative mechanism is absent in the ordinary QGSM 
\cite{DTU_DPM,DPM1,QGSM_K}. From the point of view of the TE it is a subleading 
(in $1/N$) effect. Our calculations show that its contribution
to the midrapidity net proton density at $\sqrt{s}\sim 10$ GeV 
may be quite large.\\

\noindent {\bf 4}.
In summary, we have studied, for the first time,  
the radiative energy loss contribution to the proton stopping in $AA$
collisions in the energy region $\sqrt{s}\sim 10$ GeV which
is of interest in connection with the future experiments at NICA and 
the BES program at RHIC. The analyses is based 
on the LCPI  
approach \cite{LCPI1,LCPI2}   to the induced gluon emission. Calculation are performed beyond the 
soft gluon approximation. 
We have compared our results with
the data on the proton rapidity distribution from NA49 \cite{NA49} 
for central Pb+Pb collisions at $E=40$ GeV.
We have found that  the radiative correction  can fill in partly the 
midrapidity dip in the net proton rapidity
distribution. For $m_g\sim 400$ MeV the radiative effect turns
out to be rather strong, and its contribution to the midrapidity
net proton density is comparable to the prediction of the ordinary 
QGSM. 

We argue that the net proton midrapidity fluctuations at $\sqrt{s}\sim 10$ GeV 
may be dominated by the initial fluctuations of the proton flow to
the midrapidity region. 
The fact 
the net proton
distribution from 
STAR \cite{STAR_BES} at $\sqrt{s}\sim 10$ GeV is close to the 
Poissonian/binominal
distribution provides evidence for
the diquark mechanism of the baryon flow, while disfavors the picture
where quarks (without SJ) transmit the baryon number \cite{Kovch}. 
The dominance of the initial state fluctuations makes questionable observation
of the QCD critical point via the data on the net proton fluctuations
in a rapidity window $\Delta y\sim 1$.\\

This work was partly supported by the RFBR grant 
18-02-40069mega.


\begin{thebibliography}{99}

\bibitem{RV1}
G.C. Rossi and  G. Veneziano,
Nucl. Phys. B{\bf 123}, 507 (1977).

\bibitem{Ven_TE}
G. Veneziano,
Phys.~Lett. B{\bf 52}, 220 (1974).

\bibitem{Ven_TEQCD}
G. Veneziano,
Nucl.Phys. B{\bf 117}, 519 (1976).

\bibitem{DTU_DPM}
G. Cohen-Tannoudji, A.E. Hassouni, J. Kalinowski, and R.B. Peschanski,
Phys.~Rev. D{\bf 19}, 3397 (1979).

\bibitem{DPM1}
A. Capella and J. Tran Thanh Van,
Phys.~Lett. B{\bf 114}, 450 (1982).

\bibitem{QGSM_K}
A.B. Kaidalov,
Phys. Lett. B{\bf 116}, 459 (1982).

\bibitem{KP}
A.B. Kaidalov and O.I. Piskunova,
Z.~Phys. C{\bf 30}, 145 (1986).


\bibitem{ISR}
B. Alper {\it et al.},
Nucl. Phys. B{\bf 100}, 237 (1975). 


\bibitem{KC-AA} 	
A. Capella, A. Kaidalov, A.K. Akil, C. Merino, and J. Tran Thanh Van,
Z.~Phys. C{\bf 70}, 507 (1996)
[hep-ph/9507250].

\bibitem{C-AA99}
A. Capella and C.A. Salgado,
Phys.~Rev. C{\bf 60}, 054906 (1999)
[hep-ph/9903414].

\bibitem{Z_B1}
B.Z. Kopeliovich and B.G. Zakharov,
Z.~Phys. C{\bf 43}, 241 (1989).

\bibitem{C-AA96}
A. Capella and B.Z. Kopeliovich,
Phys.~Lett. B{\bf 381}, 325 (1996)
[hep-ph/9603279].


\bibitem{Steph1}
M.A. Stephanov,
Phys.~Rev.~Lett. {\bf 102}, 032301 (2009)
[arXiv:0809.3450].


\bibitem{Asakawa2}
M. Asakawa and M. Kitazawa,
Prog.~Part.~Nucl.~Phys. {\bf 90}, 299 (2016)
[arXiv:1512.05038].


\bibitem{Asakawa1}
M. Asakawa, U.W. Heinz, and B. Muller,
Phys.~Rev.~Lett. {\bf 85}, 2072 (2000)
[hep-ph/0003169].

\bibitem{LCPI1}
B.G.~Zakharov, JETP\ Lett. {\bf 63}, 952 (1996)
[hep-ph/9607440].


\bibitem{LCPI2}
B.G.~Zakharov, JETP\ Lett. 
{\bf 65}, 615 (1997) [hep-ph/9704255].

\bibitem{LCPI3}
B.G.~Zakharov,
Phys.\ Atom.\ Nucl. {\bf 61}, 838 (1998)
[hep-ph/9807540].

\bibitem{NA49}
T. Anticic {\it et al.}  [NA49 Collaboration],
Phys.~Rev. C{\bf 83}, 014901 (2011)
[arXiv:1009.1747].


\bibitem{Ans}
M. Anselmino, E. Predazzi, S. Ekelin, S. Fredriksson, and D.B. Lichtenberg,
Rev.~Mod.~Phys. {\bf 65}, 1199 (1993).

\bibitem{D1}
V.T. Kim, Mod.~Phys.~Lett. A{\bf 3}, 909 (1988).

 	

\bibitem{D2}
C. Chen, B. El-Bennich, C.D. Roberts, S.M. Schmidt, J. Segovia, and 
S. Wan, Phys.~Rev. D{\bf 97}, 034016 (2018)
[arXiv:1711.03142].


\bibitem{SIG3}
N.N. Nikolaev, B.G. Zakharov, and V.R. Zoller,
JETP Lett. {\bf 59}, 6 (1994)
[hep-ph/9312268].

\bibitem{NPZ}
N.N. Nikolaev, G. Piller, and B.G. Zakharov,
J.~Exp.~Theor.~Phys. {\bf 81}, 851 (1995)
[hep-ph/9412344].


\bibitem{MK}
Y.V. Kovchegov, A.H. Mueller,
Nucl.~Phys. B{\bf 529}, 451 (1998)
[hep-ph/9802440].

\bibitem{BDMPS}
R.~Baier, Y.L.~Dokshitzer, A.H.~Mueller, S.~Peign\'e and D.~Schiff,
Nucl.\ Phys.\ B{\bf 483}, 291 (1997) [hep-ph/9607355].

\bibitem{Z_OA}
B.G. Zakharov,
JETP Lett. {\bf 73}, 49 (2001)
[hep-ph/0012360].

\bibitem{NZ_HERA}
N.N.~Nikolaev and B.G.~Zakharov,
Phys. Lett. B{\bf 327}, 149 (1994) 
[hep-ph/9402209].

\bibitem{DSE_m08}
Si-xue Qin, Lei Chang, Yu-xin Liu, C.D. Roberts, and D.J. Wilson,
Phys. Rev. C{\bf 84}, 042202  (2011) 
[arXiv:1108.0603].

\bibitem{DSE_m04}
J. Papavassiliou,
arXiv:1112.0174.

\bibitem{DSE_m05}
A.C. Aguilar, D. Binosi, J. Papavassiliou, and J. Rodriguez-Quintero,
Phys.~Rev. D{\bf 80}, 085018 (2009) 
[arXiv:0906.2633].

\bibitem{C-ppAA12}
A. Capella and E.G. Ferreiro,
Eur.~Phys.~J. C{\bf 72}, 1936 (2012)
[arXiv:1110.6839].

\bibitem{WNG}
A.~Bialas, M.~Bleszynski, and W.~Czyz,
Nucl. Phys. B{\bf 111}, 461 (1976).

\bibitem{KN}
D.~Kharzeev and  M.~Nardi,
Phys. Lett. B{\bf 507}, 121 (2001)
[nucl-th/0012025].


\bibitem{CK-Bflow}
G.H. Arakelian, A. Capella, A.B. Kaidalov, and Yu.M. Shabelski,
Eur.~Phys.~J. C{\bf 26}, 81 (2002)
[hep-ph/0103337].

\bibitem{NA49Lambda}
T. Anticic {\it et al.}  [NA49 Collaboration],
Phys.~Rev.~Lett. {\bf 93}, 022302 (2004)
[nucl-ex/0311024].



\bibitem{BM-entropy}
B.~M\"uller and K.~Rajagopal,
Eur. Phys. J. C{\bf 43}, 15 (2005)
[arXiv:hep-ph/0502174].

\bibitem{NA49Nch}
S.V. Afanasiev {\it et al.} [NA49 Collaboration],
Phys.~Rev. C{\bf 66}, 054902 (2002)
[nucl-ex/0205002].

\bibitem{entropy} 	
S. Borsanyi, G. Endrodi, Z. Fodor, A. Jakovac, S.D. Katz, S. Krieg, 
C. Ratti, and K.K. Szabo,
JHEP {\bf 1011}, 077 (2010)
[arXiv:1007.2580].

\bibitem{FROUT}
S.A. Bass, A. Dumitru, M. Bleicher, L. Bravina, E. Zabrodin, 
H. Stoecker, and W. Greiner,
Phys.~Rev. C{\bf 60}, 021902 (1999)
[nucl-th/9902062].


\bibitem{STAR_BES}
L. Adamczyk {\it et al.} [STAR Collaboration],
Phys.~Rev.~Lett. {\bf 112}, 032302 (2014)
[arXiv:1309.5681].



\bibitem{Kovch}
J.L. Albacete, Y.V. Kovchegov,
Nucl.~Phys. A{\bf 781}, 122 (2007)
[hep-ph/0605053].

\bibitem{LUND}
B. Andersson, G. Gustafson, G. Ingelman, and T. Sjostrand,
Phys.~Rept. {\bf 97}, 31 (1983).




\end{thebibliography}
\end{document}